# Deployment Archetypes for Cloud Applications

Anna Berenberg, Google, Inc.
Brad Calder, Google, Inc.

**Abstract**

This is a survey paper that explores six Cloud-based deployment archetypes for Cloud applications and the tradeoffs between them to achieve high availability, low end-user latency, and acceptable costs. These are (1) Zonal, (2) Regional, (3) Multi-Regional, (4) Global, (5) Hybrid, and (6) Multi-Cloud deployment archetypes. The goal is to classify cloud applications into a set of deployment archetypes and deployment models that tradeoff their needs around availability, latency, and geographical constraints with a focus on serving applications. This enables application owners to better examine the tradeoffs of each deployment model and what is needed for achieving the availability and latency goals for their application.

## 1. Introduction

In looking at how applications have changed over the past 20 years, we have evolved from a world where planned maintenance downtime was standard and business applications were typically available only 99% of the year [59] to today where applications are expected to be up and running 24/7. Similarly for latency, online transactions over the internet took on the order of seconds 20 years ago [61], where users today expect transactions to complete in milliseconds.

The drive towards higher availability and lower end-user latency is pushing application developers and operators to evolve and deploy applications with the best availability and latency possible. Even applications built around deployment options that were only available 20+ years ago need to be supported in this 24/7 available and low-latency world. With Cloud as the preferred platform for deploying and running applications, this means Cloud needs to help achieve these goals for (a) applications that have been around since before Cloud existed (Enterprise applications) and (b) greenfield applications born in the Cloud (Cloud-native applications).

As businesses move to the Cloud, some applications may need to continue to run as they did on-premises and potentially benefit from Cloud-managed services (e.g., storage, relational databases, data analytics, SAP). In comparison, some businesses may want to evolve applications within existing boundaries, or go for partial or complete rewrites to achieve higher availability, better end-user latency, and increased operational efficiency and agility. In addition, the underlying technologies used for existing applications influences the deployment options that are suitable when migrating applications to the Cloud. The choices each business will make for their applications will be different depending on the needs of each business.

When running an application in the Cloud, there are many important aspects an application owner needs to address including security, identity, data recovery, data and traffic management, cost optimization, and much more. We touch on some of these, but this paper is primarily focused on exploring different deployment models for the application serving stack.

We explore six Cloud-based deployment archetypes for Cloud applications and the tradeoffs between them to achieve high availability and low end-user latency. These are (1) Zonal, (2) Regional, (3) Multi-Regional, (4) Global, (5) Hybrid, and (6) Multi-Cloud deployment archetypes. Cloud applications consist of multiple services and microservices, and the application as a whole may mix services of different archetypes based upon its needs. We look at multiple categories of application deployments from Enterprise to Cloud-native applications, their impact on availability and latency, and how they can leverage these six deployment archetypes.

## 1.1 Principles of Availability

The level of availability each part of the application is targeting depends on its business purpose[14]. Some applications only need 3 nines (99.9%) availability, which means the service can be unavailable for at most 43 minutes a month. Other applications need 4 nines (99.99%) availability, which means the application can only be unavailable for at most 52 minutes a year. Then there are those mission-critical applications that need 5 nines (99.999%) availability, where they can only be unavailable at most 5 minutes a year. To achieve these levels of availability, it is important to understand what is needed for each part of the application, and invest in closing the gap between current and desired availability for each part.

The investment in availability comes at a cost, but it is often crucial to the long term success of the business, since availability directly influences the reputation of the business and the satisfaction of the application's users. For the purpose of this paper, we group together into the **overall availability of the application** the following: (a) the time to access the application, (b) the time to get a response with valid results, (c) the application's access to its data, (d) the assurance that data is stored and maintained with integrity, and (e) the application's ability to scale and handle peak traffic demands.

For an application, availability is best designed from the start. Adding availability as a feature later on can require re-architecting the application and potentially a full rewrite. A key part of the design is how the application reasons about fault domains and how it provides redundancy and scales across those fault domains to maximize availability. A fault domain is a set of infrastructure parts that together represent a single point of failure. To increase availability, applications need to run and store their data across multiple fault domains (zones and regions) and have the ability to balance load or fail over in case of failure. Data needs to be replicated and backed up so that it is never lost, and checks must be in place to make sure data is never corrupted. In addition, applications need to be able to quickly load balance across multiple instances of the application to scale to the largest traffic the application can have. This includes minimizing time for startup and shutdown, so applications can be restarted, and scaled up and out quickly.

Two additional important concepts for minimizing the impact of an outage are (a) sharding the application and (b) making sure all application updates are done incrementally and can be rolled back. Applications may apply sharding across their users or data so that they are served across different fault domains [2]. In this way, an issue with one fault domain will only impact a subset of the users/data, thus containing the failure radius (often called the blast radius) [33]. Similarly, code and configuration changes should be rolled out incrementally across the different fault domains to gradually introduce a change into production, with the ability to quickly roll back if any production issues are discovered in order to return the application to a healthy state. This allows code and configuration production issues to be discovered early on, and reduces the impact to only those parts of the application running in the fault domains being updated. In addition to being able to quickly roll back recent application changes, having the ability to drain or shed load from the affected fault domains is often used to quickly mitigate issues.

These techniques collectively determine how large of an impact there is to the application and its users when there is an outage. Ideally there is no impact on users when an issue occurs, but if the best design and deployment practices are followed, when there is an issue then only a small set of users of the application are affected in one or a few fault domains.

Finally, applications need to understand their dependencies, the availability and failure modes of those dependencies, and to evaluate the multiplicative implications across these dependencies on the application's design and availability. Typically, the fewer dependencies a service has the better, and it's better to avoid linking in code, calling out to other services and APIs that bring in unknown dependencies. As part of the overall manageability, separating out parts of the application into its vital and non-vital services, identifying the availability targets of each, and continuously improving the vital parts, are important. If a service is vital, then for its vital components, all of their dependencies (recursively down the call chain) should be either highly available or the component should be able to function in the absence of the dependencies. Examine availability for each service in the application independently and for the application as a whole.

In this paper, as we examine the different deployment archetypes, we examine the availability applications can achieve with each archetype with the focus on overall availability as described in this section.

### 1.2 Types of Applications

A business relies on multiple types of applications, each having different availability and latency requirements.

- **Business-critical applications** - these represent the critical applications for a business. If these applications are unavailable, the business is down. Highest availability and lowest latency are desired for these applications. These applications could be user-facing or not, and the classification of a business-critical application depends on each business.
- **Line-of-business applications** - these are applications that support running the business. While these applications do not serve live traffic, they are typically instrumental for supplying data for the business. They often have requirements to finish work by a particular time. Continuous Integration and Continuous Deployment (CI/CD) pipelines fall into this category, as well as data processing and analytics. High availability is desirable for these applications, but they can sustain short-lived outages without having an immediate business impact.
- **Internal applications** - these are applications that are for internal consumption for a business (e.g., recruiting, time-off tracking). Best effort availability is required. Employees want these to be always available, but if they are not, the impact on the business is lower.

This paper is primarily focused on business-critical applications, though the deployment archetypes can also apply to the other types of applications as well.

### 1.3 Data Durability, Availability and Backup

For an application to be available, its data has to be available, and for the application deployments we examine there are several concepts that are related to how the data is stored and managed. These include:

- **Data durability** - long-term data protection, where the stored data does not suffer from corruption and is not lost or compromised. To achieve this, the underlying data storage system often replicates the data and performs error-correcting checks and scrubbing of the data to prevent data decay.
- **Data availability** - access to data upon request. High data availability is achieved by placing and replicating the data across more than one failure domain, keeping the data durable, ensuring the service provides access to the data, and making sure the data is appropriately resource-provisioned to serve requests. The type of replication, asynchronous (eventually consistent) or synchronous (strongly consistent), along with data failover capabilities are important building blocks for achieving data availability.
- **Data backup** - point-in-time snapshots of data. Backups are important for protecting against application and human errors, and can also be used as a means for disaster recovery. All applications should use backup services to protect against accidental loss or corruption of data due to application-level issues.

Similar to defining the desired level of availability, it is as important to define objectives for disaster recovery [64] for each application, specifically Recovery Point Objective (RPO) and Recovery Time Objective (RTO), and to choose deployment models and infrastructure that can achieve the desired RPO and RTO. RPO captures how old a copy (backup or replica) of the data is compared to the current state in production. It is important to understand the RPO, since this copy of the data would become the active state of production in case the current state fails, and any data changes more recent than the recovery point would be lost. If an application has more than one source of data, then independent recovery points of each one need to be reconciled. RTO captures how long it takes to restore an application during recovery to bring it back online and available in production, which includes access to the data needed for running the application. A highly available application wants both RPO and RTO to be as close to zero as possible.

In this paper, we examine deployment archetypes and deployment models within each archetype that focus on the availability of data, while maintaining durability, and Google Cloud [8, 27], Microsoft Azure [5, 45 ] and AWS [7, 11] have each developed storage and database products that meet the requirements needed for each deployment archetype examined. In addition, data backup services should be used in conjunction with these deployment options.

**1.4 Six Deployment Archetypes for Cloud Applications**

The following is an overview of the deployment archetypes we examine in this paper:

1. **Zonal** - All components of an application run within a single zone. A zone provides a set of clusters with the infrastructure needed to run services (compute, storage, networking, data, etc) within that zone. Should a zone go down, what is running within the zone is either restarted in another zone from the last checkpointed state, or a failover occurs to a standby instance of the application in another zone.

2. **Regional** - All components of an application are deployed and run out of one Cloud region. A region consists of 3 or more zones, where each zone is treated as a separate fault domain. High availability can be achieved by replicating the application across zones within the region. These applications are typically designed to run with a data store that shares data and makes it accessible across that region. To serve application traffic, the requests are load-balanced across compute instances in multiple zones. To further increase availability and reliability, some applications may have a secondary standby region with an asynchronous copy of the data, where the application can fail over to the secondary region in case the primary region is not available.

3. **Multi-Regional** - The application serving stack runs and is stitched together across multiple regions to achieve higher availability and low end-user latency through geographic distribution. In this deployment archetype, data is typically replicated and shared across regions. This archetype is commonly used for applications that want to achieve high availability, such as user-facing applications.

4. **Global** - The application stack is spread and replicated across Cloud regions around the globe and data is available worldwide via global databases and storage. Applications consisting of a large number of services and microservices benefit from this deployment archetype. This is the five-nines deployment model used by retail, social media and other businesses requiring always-on availability, while running large services economically.

5. **Hybrid** - Applications that have deployments combining on-premises and public cloud(s) are becoming increasingly common. On-premise software stacks will continue to evolve and be connected with the Cloud, to the point where on-premises will be considered to be another form of connected zone or region. Hybrid application availability and resilience is often achieved by (a) creating deployment archetypes that leverage failover between on-premise and Cloud, and (b) coordinating the execution of parts of the application that run in the Cloud versus run on-premises.

6. **Multi-Cloud** - Applications can potentially gain the highest availability by using two or more public Cloud platforms at the same time, to protect against one Cloud's unavailability. In each cloud, one of the deployment archetypes listed above is used, and then combined across clouds to create a multi-cloud deployment. This deployment archetype is in its infancy, but applications that require the highest availability are prime targets for multi-cloud deployments as this model evolves.

There are many reasons why one archetype will be used for an application over another. For applications that are required to have data reside in a particular region or jurisdiction, the choice of geographical distribution may be limited to an individual country or to a union of countries, and therefore the choice of deployment options will be limited. Other globally ubiquitous applications may have latency budgets, where, if latency is too high, users may interpret it as an availability issue and abandon their requests.

Within each archetype there are multiple models that represent the deployment scenarios applications may use. We will now examine each of these deployment archetypes and models, and their tradeoffs in detail, and conclude with a summary comparing the tradeoffs.

**2. Zonal**

In the Cloud, a zone represents a fault domain in which to deploy and to run services and infrastructure. Running an application within a single zone typically means running the application within a compute cluster potentially spread across multiple racks near each other in the same datacenter. Should a zone go down, what is running within the zone is either restarted in another zone from the last checkpointed state, or a failover occurs to a standby instance of the application in another zone. We now go through these two types of zonal deployment models.

## 2.1 Single Zone

Running an application only within a single zone is not targeted towards high availability, since a zone is considered as a single failure domain from both software issues as well as other types of disasters (e.g., fire). Even so, applications that need supercomputer-like connectivity, as well as applications that do not need high availability, leverage single-zone deployments.

High Performance Computing [37] and TPU Pods [28] are examples of Cloud applications that are deployed and run in a single zone. These applications typically require very low latency and high bandwidth usage, achievable within a single zone. They do not serve live traffic and can work with 3-nines availability, and they can restart from the last checkpointed state. The data for these applications can be kept in a regional data store, with the primary or one of the copies of the data stored within the zone where the data is being read and processed. In addition, an advantage of keeping applications that have a lot of communication across VMs within the same zone is that Cloud providers typically have an additional charge for egress between VMs across zones.

Another important application type that works well with single zone deployments is developer testing workloads. This enables developers to continuously build and test their applications in the Cloud. It also may be suitable for use cases where downtime is acceptable or the application can be restarted elsewhere.

A single-zone application should be considered sufficient for these use cases, but not for most production applications.

## 2.2 Primary Zone with Failover Zone

As companies bring their on-premises applications to the cloud, a first step often taken is to choose a deployment model to run the application in the Cloud with minimal changes. Some of these may be commercial off the shelf (COTS) applications that application owners acquired and may not be able to change. In addition, sometimes these applications come with per-instance licenses that can be prohibitively expensive to deploy for redundant extra copies. As a result, single-zone deployments continue to be a valid deployment option for these applications.

Single-zone applications still need as much redundancy and availability as possible. The deployment model typically used is to run the application in a primary zone, and to use a failover zone in the same region as a recovery zone. If the primary zone has issues, the recovery zone is used to start the application up again. Many enterprise applications are built to run in a form of primary/failover configuration, also known as Highly-Available (HA) topologies, and this is an established pattern used in enterprise and on-premises deployments over the years.

Let's look at the example of a single-license application running in the Cloud that wants to have failover support. Assume there are two VMs in two different zones (A and B), where one is the primary and the other is used as the failover. In this example the application owner has to pay for every instance running, so the application is only running in the primary zone and not the failover zone to control costs. In this case there are generally three options for how to connect to the application VM for license renewal:

- **Static IP address** (also referred to as floating IP address) - This static IP is used for the license renewal and can be either private RFC 1918 [54], RFC 6598 [62], or public IP. The static IP address initially points to the primary VM (for this example assume it is in zone A), which runs the single application. When zone A goes down, either manual or script-based reconfiguration occurs, and the application is started in zone B, with the same static IP address. In this case clients can continue to connect to the same IP address, whether they use DNS resolution or connect directly to the IP.
- **List of Static IP addresses** - List of IP addresses is used in a round-robin fashion in case connection is lost. The exact logic to pick one address from the list depends on application client-side behavior.
- **Dynamic addresses with DNS** - If the IP for license renewal is not static, then DNS is used for resolution. In this case, DNS is configured to point to the primary VM in zone A. When zone A goes down, the DNS configuration is updated to point to the VM in zone B. The tradeoffs around DNS and how it relates to failover deployments are discussed in Section 4.2.

Now let's look at another example in Figure 1, which is a basic application deployed in a primary zone with a replica for failover purposes in a secondary zone. In this example we have a Load Balancer (LB), which denotes not just one instance, but a highly available replicated setup. The setup has a replicated compute workload named "Front-End", and the Cloud-managed database that holds the application data replicated across zones. Most databases will work in this configuration, and for this example we assume it is a SQL database.

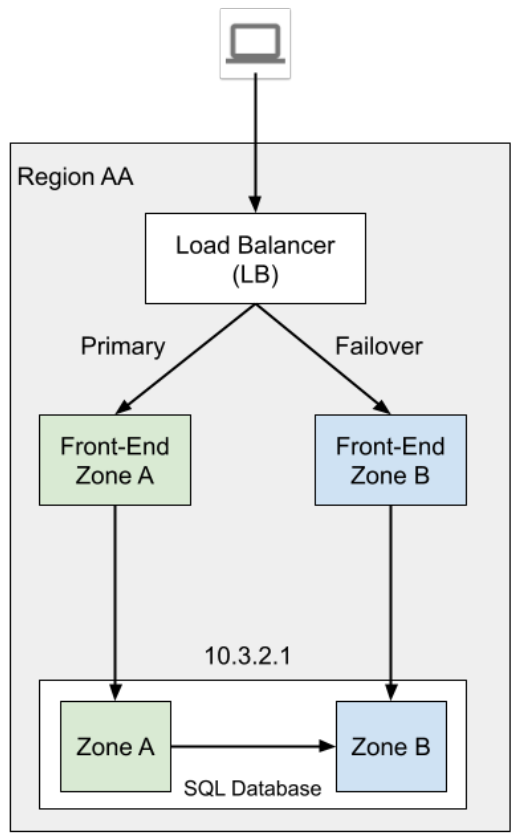

Figure 1: Single zone with failover zone deployment model.

Let's consider zone A of region AA to be a primary zone, and zone B of region AA a failover zone. The primary instance of the SQL database is placed in zone A and all read and writes happen to this instance. In addition, the SQL database is configured with a standby in zone B, and the data is replicated from zone A to zone B by the cloud provider managed database. The Front-End in zone A and Front-End in zone B are configured identically with the same virtual IP address (10.3.2.1) to access the SQL database. This means the Front-End service does not need to change the IP address of the SQL instance when failover occurs. In addition, the load balancer is configured to have a primary set of compute instances (VMs or containers) in zone A and failover instances in B.

Now let us assume that zone A fails. Every second, the primary Front-End and SQL instance in zone A responds to a heartbeat signal from the monitoring system. If multiple heartbeats are not detected by the monitoring system, an alarm is sent and failover is initiated by the application owner or a script that has been automated. With failover initiated, the Front-End in zone B now serves user traffic, and the standby SQL instance in zone B is configured to now act as the primary SQL instance using the same virtual IP address (10.3.2.1) [22]. The load balancer will react to the failure in zone A by moving traffic to zone B, because it was configured to fail over to the Front-End in the other zone based on health check status. Once the traffic is being served from zone B and the primary SQL instance is now in zone B, the availability is re-established for a single zone application on failover.

Health checking is an essential part of the failover process. As part of the health check status, the application's services need to decide if they are healthy or not, and this greatly depends on the service. Each instance of the service needs to determine its health based on error rates, exhaustion of resources such as CPU and memory or other custom signals, and declare itself unhealthy as part of a health checking response.

When zone A comes back, the traffic is not sent back to zone A by the load balancer unless the application owner decides to fail back. The deployment will now be in a steady state with zone B as primary and zone A as failover, until a failover is performed to make zone A the primary again. A best practice here is to reserve the capacity needed for failover in the failover zone and ready to go in case of a failure, and to routinely fail over the application between zones to ensure failover works when it is needed. Note, there are additional scenarios to cover for an application using failover (e.g., restarting up cross-zone data replication after failover, whether to allow failing over only part of the application stack, and more), which we did not have time to go over in this paper, and the application owner needs to make sure they are covered for this type of deployment model.

**3. Regional**

In Cloud, a region is a specific geographical location in which to deploy and run application services and infrastructure that consists of multiple zones. A region consists of three or more zones, where each zone is treated as a separate fault domain. High availability can be achieved by replicating the application and its data across zones within the region.

We distinguish between zonal and regional archetypes with the following definition. The regional archetype has the application replicated across multiple zones within the region and actively serving traffic across the multiple zones at the same time. In comparison, the zonal archetype has an application serving traffic from a single zone and then failing over to another zone when there is an issue.

Single-region applications typically focus on users in one geography (e.g., country). This is used to (a) optimize for latency, where users are served from the same region they reside in, and/or (b) provide data location requirements, where user data is kept and served from a single country or region. To further increase availability and reliability, some applications may have a secondary standby region with an asynchronous copy of the data, where the application can fail over to the secondary region in case the primary region is not available. We now describe these two deployment models (single region and a single primary region with failover).

**3.1 Single Region**

In the context of this paper, running an application in a region means running an application spread across multiple zones within that region, where each zone is treated as an independent failure domain. A best practice here is to replicate the application across all of the zones within the region and keep the size of each deployment approximately the same across zones. This ensures the application always has capacity available in other zones when there is a zonal failure.

To demonstrate this, we will use a more complex application architecture shown in Figure 2. In this example, we have a service named "Front-End" that contains the interface the end-users interact through, a service "Back-End" that contains the business logic of the application, and a Cloud-managed database (e.g., SQL) that holds the application's data. In addition, there are load balancers in front of each service to load-balance requests across them. A request flow across the services in the diagram can be described as *user->Front-End->Back-End->SQL* and the response in the opposite order. In reality, applications consist of a large number of services and microservices, ranging from tens to hundreds in the same application.

As shown in the example, the single-region application should try to achieve higher availability by replicating data as well as compute workloads across multiple zones within the region. To achieve this for data, most Clouds support replicating SQL synchronously across the zones within a region, where writes and reads go to the primary zone for the SQL instance even though the data is replicated across zones for durability [24]. In addition, there can also be read replicas configured across all zones, where read requests are served from the closest zone.

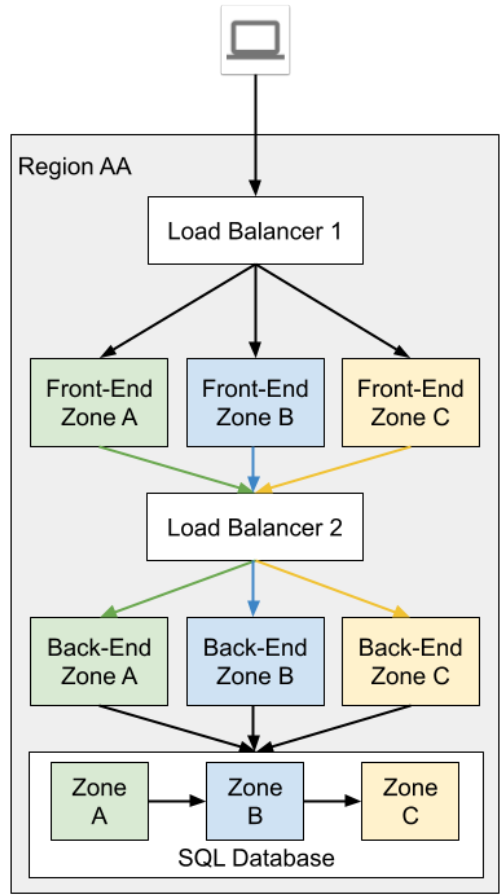

Figure 2: Single region deployment model.

When a request comes into Load Balancer 1, the request may be forwarded to any healthy replica of the Front-End service in any of the zones in the region, and then the request path is latency-optimized to keep the request flow within the same zone as the request traverses multiple services. For example, a request arriving at Front-End(A) will be sent to Back-End(A) to optimize latency by keeping the request within the same zone. Then the request to the SQL database may be across zones. The SQL(primary) is running in one of the zones, and communication with the SQL(primary) may happen across the zones, as writes are always sent to the primary.

Should one zone become unhealthy due to software failure, such as bad binary rollout for the Front-End or Back-End service, or due to infrastructure failure such as a power outage event, the requests destined to that zone will be steered to another zone. If the zone where the primary instance of SQL fails, automatic failover is initiated [23] for the SQL(primary) and a standby replica now becomes the new primary. To achieve availability during failover, applications should retry idempotent requests or re-establish a new connection to the database [60].

As an example, assume that the SQL(primary) is in zone B and assume the Back-End microservice in zone B becomes unhealthy. For a request that comes into Front-End(B), the Load Balancer 2 will know that Back-End(B) is unhealthy and choose a different back-end to route the traffic to. In this case, the traffic flow coming into zone B is now *Front-End(B)->Back-End(C)->SQL(B) or Front-End(B)->Back-End(A)->SQL(B)*. With this approach the load balancers can route the traffic around failures within a specific zone at a specific service layer.

For a single-region deployment, a question that comes up is how many zones to run the application across. The standard topology in Cloud is to have three zones for each region, where an application is run across all 3 zones and it uses the Cloud provider managed regional storage and database services to manage the data. The reason for using 3 zones instead of 2 zones is because the loss of one zone means losing 1/3rd of the serving capacity instead of 1/2 of the capacity. Not having enough serving capacity left after the loss of the zone can impact availability as it takes time for autoscaling to kick in. Another reason is to survive the unlikely but possible event of two simultaneous failures across two zones - one caused by the application and the other by the Cloud provider. This can potentially occur if (a) an issue in the application causes a single zone outage (e.g., as the application is incrementally updated one zone at a time and an outage is potentially found after the first zone is updated) and (b) the Cloud provider has an issue that results in one zone having an outage. With 3 zones, an application can still be available even if two zones have issues (one due to the application update and one due to the Cloud provider).

**3.2 Primary Region with Failover Region**

While single-region applications with multi-zone replication provide a highly available region, some business applications may have continuity requirements over large distances (e.g., having a primary and secondary separated by hundreds of miles). The desire for business continuity for a single-region application is fulfilled by maintaining a second region that is used for failover events. To satisfy compliance requirements, primary and standby regions may need to be located in the same country or union of countries. If there is no compliance requirement, the failover region may be located anywhere where the latency increase on failover for serving response time is satisfactory.

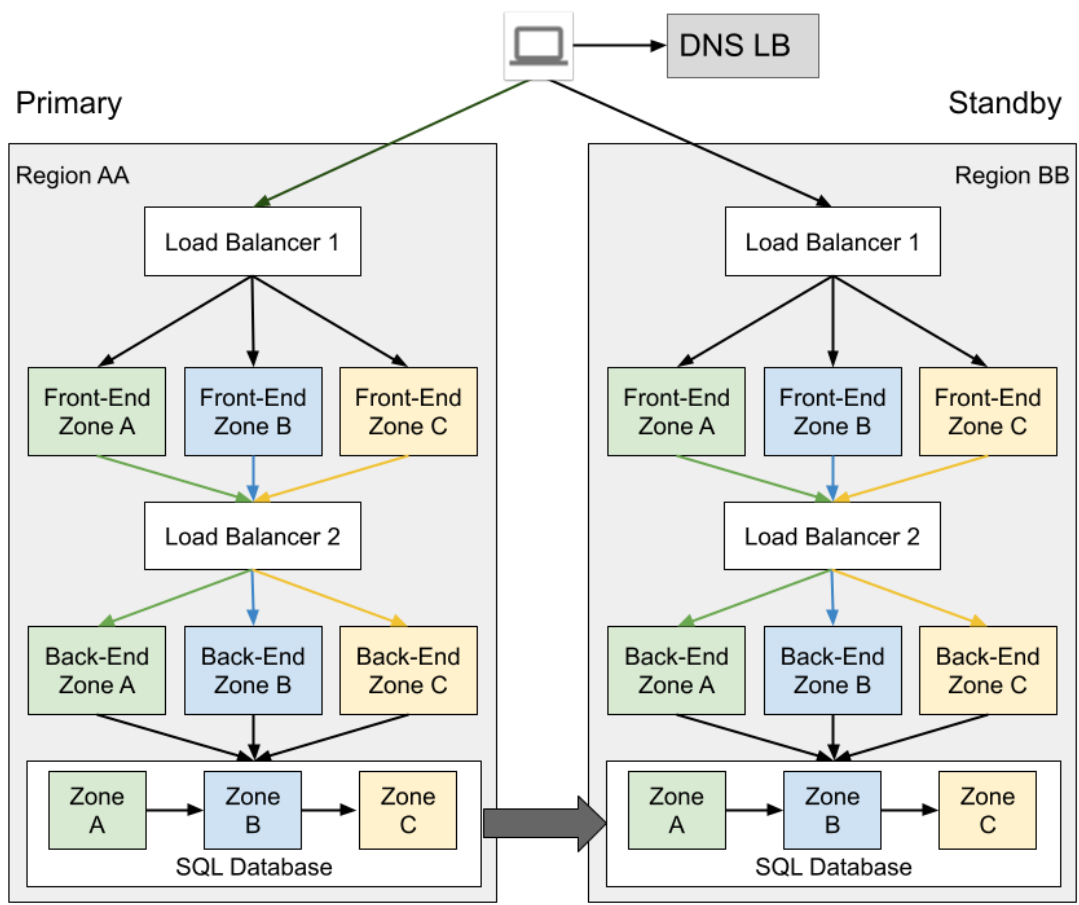

Figure 3: Primary region with failover region deployment model.

In this deployment model as depicted in Figure 3, application data is synchronously replicated within a primary region, providing RPO=0 for in-region failures. It is also asynchronously replicated to a standby region [21] that is sufficiently distant from the primary region. While this means a non-zero Recovery Point Objective (RPO) and therefore potential data loss of recent updates on failover, the approach is used by Enterprise applications with availability or regulatory needs that require a replica in another region. Live traffic is always served from the primary region, and if the primary region becomes unhealthy either due to infrastructure or software component problems, then the standby region is used.

Some application owners prefer manual failover to the standby region. In this scenario, the DNS entry for the primary region is manually substituted with the VIP or IPs of the standby region when failover occurs. Otherwise, DNS Load Balancing (DNS LB) is used for automatic failover, which we describe in Section 4.2. If DNS is not used at all, then clients have the list of IPs for both the primary and standby region and they are configured to use the current primary region.

For this model, the deployment is regional aside from the DNS LB. The DNS LB assigns traffic to the primary region, but if there is an issue with the primary region and a failover needs to occur then DNS LB assigns traffic to the standby region. Health checking is done by the DNS LB sending probes to load balancers that represent a region (the Load Balancer 1 in Figure 3). If the health checks fail then the application can fail over to the standby for availability. Note, the DNS LB can also be used on an application owner's Virtual Private Cloud (VPC) for service to service communication as a service discovery mechanism.

For an application with more than one region to be operational, there needs to be an understanding of the full health of the service stack within a region. If there is a regional issue for just a single layer (e.g., Load Balancer 1, all Front-Ends, Load Balancer 2, all Back-Ends, or regional SQL) then the region would be unhealthy and a failover would need to occur to keep the service up and running. This means the application owner needs to build up an understanding of the health of all layers and be able to trigger failover if a given layer is having a regional issue. We will go into more details on this in Section 4.3.

When a failover is triggered, the primary for the database is switched to the new primary region (what was the standby database will be promoted to be the new primary [25]). If this is an unplanned failover then recent updates to the database could be lost. For a planned failover, the failover can be coordinated to ensure all of the latest changes from the primary are made to the standby database before switching over.

With this deployment model, there is a need to ensure that the standby region functions well even though the majority of times the region is idle. The best practice is to perform planned failovers on a timeline that makes sense for the business. In addition, health probers should be used to not only continuously check the health of the primary region, but also the standby region.

For some applications, this deployment can be simplified to have a primary region with a single zone and failover region with a single zone. This targets applications that are limited by number of licenses or limitations in architecture, and is an improvement over the Primary Zone with Failover Zone deployment described in Section 2.2 for applications that need cross-regional business continuity.

## 4. Multi-Regional

The world has become more and more interconnected, with the users of the applications becoming more and more geographically dispersed. With that, the application deployment that was traditionally optimized for availability needs to evolve because user-perceived latency has become a differentiating factor between competing applications. In addition, as users moved from desktop to mobile they have gotten accustomed to having access to their application and data anywhere with quick response times. This has pushed application deployment to being able to serve requests near to where users reside, and means a single-region application is becoming less competitive from a user latency perspective.

Running an application across multiple regions gives lower end-user latency, drives higher availability, and meets some business continuity requirements by having the application and its data running and available in multiple regions separated by hundreds of miles. There are different deployment models for multi-regional archetypes, and we'll go through two of them.

### 4.1 Fully Isolated Stacks with Data Sharding

If the application data is partitionable into separate databases, then one deployment option some applications have considered is to partition or shard the application and data across multiple regions into separate isolated stacks. In this approach, each stack would use the Primary Region with Failover Region approach described in Section 3.2, and a given user's data is confined to a single regional deployment based on the sharding.

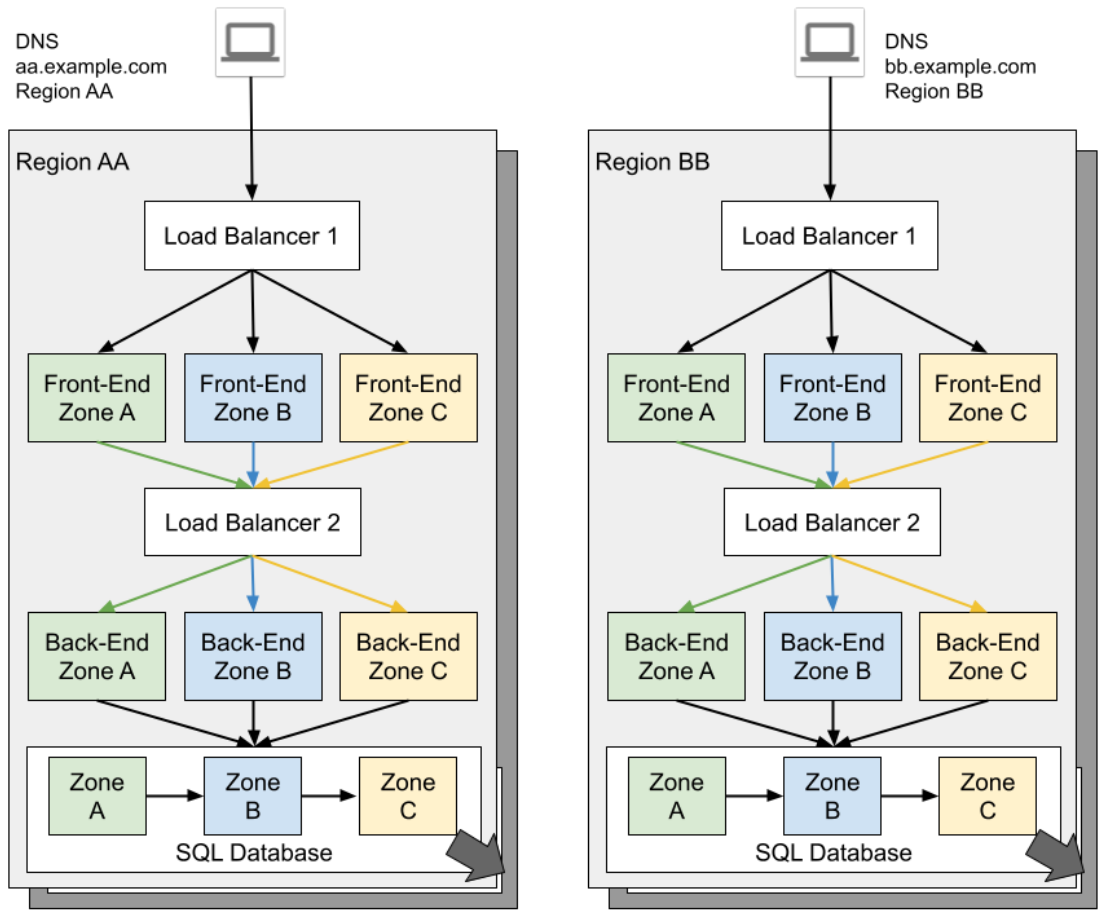

Figure 4: Fully isolated stacks with Data Sharding in multi-regional deployment model.

This is shown in Figure 4, where users are routed to the region where their data resides, and their requests are fully processed within that region. A given client of the application knows the region that it is accessing and requests an application name resolution by regional hostname - our example application in region AA has a hostname aa.example.com and in region BB it has a hostname bb.example.com. We do not imply geographical proximity between the regions AA and BB. In practice two regions could be close to each other (e.g., 10ms to 20ms Round Trip Time) or far away across continents (e.g., 100ms to 200ms Round Trip Time).

For applications that can only be run within a single region, this gives potentially higher availability by sharding user data across multiple regions and better end-user latency for the users close to the regions hosting their data. Sharding the user data across regions can reduce the number of users impacted due to an application change that affects just a single region. In addition, this deployment model can be used to meet jurisdictional requirements for keeping data within a given region. This approach has the disadvantage of (a) having to deal with failover and loss of availability when a single region has issues, (b) having issues absorbing distributed denial of service (DDoS) attacks and large traffic spikes being directed at a specific region, since user requests go to a specific region and cannot be load-balanced across multiple regions, and (c) rendering user experience as a function of where their data is located (i.e., if a user's data resides in region AA then it is always routed to region AA no matter where the user is in the world).

**4.2 DNS Load Balancing**

The next step in the evolution of multi-regional applications is the addition of a DNS Load Balancer (LB) to connect regional application stacks and load balance traffic across the regions. This is for applications that can run across multiple regions and be run with a data store that shares data and makes it accessible across those regions.

DNS-based Load Balancing [4] is considered to be the standard way for clients to resolve the domain name of the service to get the IP address to use when accessing the service [4, 43]. A domain can be configured with one or more IP addresses, usually Virtual IPs (VIPs), and these VIP addresses target load balancers that front the application stacks. Then the DNS is configured by the application owner with one of the following routing policies that determines how the VIP addresses are given out for client requests:

- **Round robin (RR)** - DNS requests are rotated and shared evenly across multiple IPs/VIPs that serve a domain.
- **Weighted round robin (WRR)** - DNS requests are assigned to different VIPs based on service owner-configured weight.
- **Geo-mapping of clients** - Another option is to create geo-mapping of clients to an edge region and DNS requests will be assigned to the closest IP/VIP. For this approach, DNS LB providers have their own knowledge of IP prefixes mapped to known geographies, as well as latency associated with reaching these geographies. This mapping is used to decide which region client IPs belong to, as well as which region destination VIPs belong to when processing a request.

When DNS is integrated with load balancing, the load balancer health-checks the application VIPs by sending requests to the VIP as if it were with real traffic. Typically such health checking is done from several regions to make sure there is a reasonable level of confidence that the VIP is indeed healthy. After the health status is collected, unhealthy VIPs are removed from participating in the DNS routing assignments. In addition, DNS load balancing can be set up so that DNS requests can be redirected from a primary VIP to a failover VIP based on health checking.

With DNS, a client initiates a DNS request every time when the DNS' time-to-live (TTL) expires. This resolved address is the address the client will send the request to. The client will continue to use that resolved address until the TTL expires and continue to send requests to the assigned region and its services behind the domain. The TTL affects the failover time and availability of the application, and ultimately the user's experience.

We now walk through an example using Figure 5 with region AA and region BB. When a request to resolve example.com is answered by the DNS LB it is mapped to a VIP based on which one of the 3 routing approaches described earlier is used. If region AA becomes unhealthy and the DNS LB knows this, then it will update the DNS routing to not use AA. For clients already using AA, they will need to wait for their TTL to expire before they contact DNS again; then they will be routed to BB. The TTL will affect the availability of those clients during this resolution.

We now look at the tradeoff of using DNS LB with geo-mapping versus weighted round robin. Let's assume 100 DNS requests come from clients near the geographical area around region AA and 50 DNS requests come from clients near region BB.

- With DNS geo-mapping, region AA will receive all traffic generated by clients from region AA, and region BB will receive all traffic generated by clients from region BB. This produces the best latency, but an unequal load on the two regions, and can lead to resource contention in some regions over others. In this case, it can be beneficial to use autoscaling to scale each region to the desired load.
- With DNS weighted round robin, that says to send 50% of requests to region AA and 50% to region BB, this will cause 75 DNS requests to go to region AA and BB each. This provides the best load distribution, but latency will suffer because clients will be mapped across regions randomly. In addition, even for distributing load this is an approximation, since DNS does not understand the actual load received by the services in each region, and is instead assuming the load will even out over time. But this may not hold true, since a given client may generate significantly more requests for example.com than other clients. Therefore, the number of requests sent to a region do not directly correlate with the number of DNS requests.

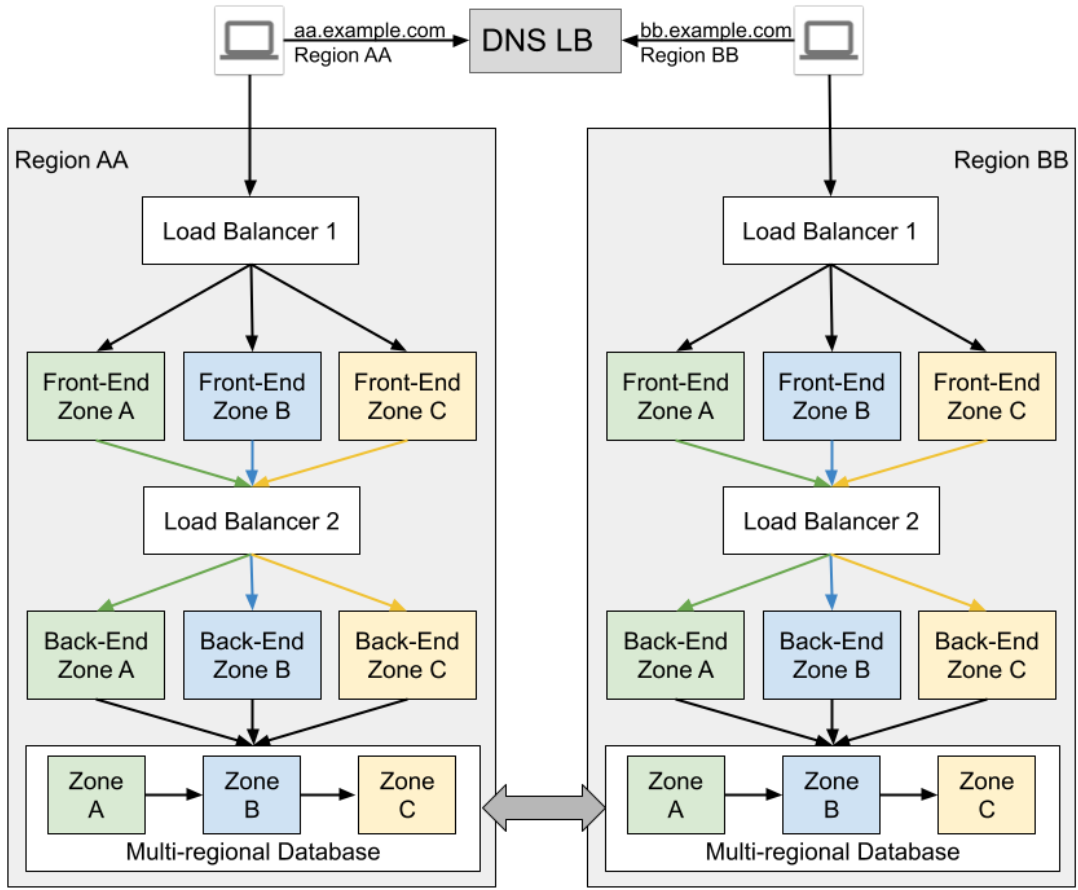

Figure 5: DNS with multi-regional isolated stacks deployment model.

### 4.3 DNS Load Balancing with Isolated Stacks

With DNS load balancing, user data should no longer be contained to one region for primary operations, since requests can end up being sent to any of the regions participating in the DNS load balancing. For this deployment model, multi-regional or global storage and database solutions should be used, since the same data needs to be accessible at the same time across multiple regions. Asynchronous or synchronous cross-region replication is employed depending on capabilities of storage and databases as well as the need of the application [8].

In order for DNS load balancing to truly work and provide the highest availability, there needs to be an understanding of the end-to-end health of the service stack within each region, and it is **up to the application, or the monitoring infrastructure, to combine the end-to-end health of the application together and provide this to the DNS Load Balancer**. In this Section we will look at two strategies for providing full-stack health information to the DNS LB.

For multi-regional topology, the DNS Load Balancer usually is configured for health checking, combined with load balancing or routing options. For each regional VIP there are one or more VIPs dedicated as backup pointing to the other regions. If a health check of the regional VIP fails, backup VIPs are used. If there is more than one backup VIP, then configured load balancing policies are used (e.g., RR, WRR or geo-mapping) over VIPs in the backup pool.

An alternative approach to having backup VIPs is to have DNS with the set of regional VIPs and weights used to guide the traffic across those VIPs. Health checks can be attached to load balancing policies and an unhealthy VIP is removed from the VIP pool; in the WRR case the dynamic weights are recalculated to send traffic in proportion to the remaining VIPs using the configured weights. Let's say the intended configuration is VIP-AA:0.5, VIP-BB:0.25,

VIP-CC: 0.25, which means 50% of DNS responses contain VIP-AA, 25% of responses contain VIP-BB and 25% of responses VIP-CC. If VIP-AA becomes unhealthy, DNS LB recalculates weights as VIP-BB:0.5 and VIP-CC:0.5, which means 50% of DNS responses contain VIP-BB and 50% responses contain VIP-CC.

Let's look again at Figure 5 where we are using DNS with multi-regional isolated stacks. If there is an issue for just a single layer (e.g., Load Balancer 1, all Front-Ends, Load Balancer 2, all Back-Ends, or regional SQL database) in a single region, then the region would be unhealthy and traffic should be directed to a different region. This means the monitoring infrastructure needs to build up an understanding of the health of all layers for each region and provide this to the DNS LB. There are two main approaches for achieving this:

- **Propagate layer failure up the stack to Load Balancer 1** - This approach assumes that the health of a layer is continuously propagated up the stack to each prior layer. In case of a failure in the region all Front-Ends will eventually know that they are unhealthy, which tells Load Balancer 1 there are no Front-Ends to send the traffic to in the region. This will cause DNS LB health checks to Load Balancer 1 to fail and the region to stop being used. For this approach to work, every service in the application stack needs to implement such logic.
- **Aggregate health of all layers and report health to Load Balancer 1** - The stack of services within a region is declaratively defined and a region is considered healthy only when all services are healthy via aggregated status that is collected by an independent health observer service for the application. The observer aggregates the health status across all services in the stack and sends the combined health status to Load Balancer 1. If the aggregated status is unhealthy, then the DNS LB will fail a health check and take the region out of service.

For both of these approaches, Load Balancer 1 can only front one domain name at a time (e.g., example.com), and there are usually multiple services deployed behind one domain name (e.g., example.com/videoshare, example.com/news, example.com/shopping), separated by the path or other routing attributes. The issue is example.com has a set of VIPs shared across all of these services, so there is not a way to distinguish between the different services (paths) at the DNS load balancer, which would be required to understand the health of each service and act upon it. This means to understand and give the health status to Load Balancer 1, the health of all services behind that domain name need to be combined and will have the same shared fate if there is a failure. For example, if example.com/videoshare has an issue in a region and is down, the aggregated health check will fail for example.com telling Load Balancer 1 to not use that region for any of the services under example.com, and all requests will be sent to other healthy regions. This aggregated health check creates a shared fate for services under a single domain name when using DNS load balancing for a region failure.

To summarize, the advantages of DNS load balancing with separated stacks are**:**

- Any region can serve a user request. This allows (a) a potential DDoS to be mitigated by a larger pool of resources across multiple regions and (b) traffic to be shifted to the remaining available regions if there is a failure with a specific region.
- The service owner has manual controls over per-region traffic distribution via the DNS LB configuration.
- Separate VIPs behind DNS can easily point to completely different deployments, different Clouds or on-premises data centers, providing mix-and-match options for increased failure isolation.

The disadvantages are:

- The need to implement propagation or aggregation of health across services and zones in regional stacks introduces considerable complexity.
- DNS TTL delays actuating failover to a healthy region and the time to failover is not deterministic. 75% of the TTLs are at 5 minutes and the remaining 25% are longer (can be hours to days) [10]. In addition, DNS TTLs can be ignored by some nameservers and some clients, which means the TTL can be much longer for clients than what the service provider specifies.
- DNS load balancing is based on DNS requests that do not represent volume of actual traffic and therefore cannot anticipate how much regional capacity will be needed to serve traffic represented by these DNS requests.

- Application capacity can be stranded within a region, especially for applications that are not autoscaled and applications that have diurnal traffic patterns. Capacity is not available to other regions for use since the DNS LB may not have the means to take into account the region's capacity when directing traffic.

This deployment architecture has been a standard for user-facing applications for many years.

**4.4 DNS LB with Custom Multi-Regional Load Balancing**

Large companies, such as Netflix [17, 35], who choose a multi-regional deployment but would like to build globally ubiquitous applications supplement the Cloud provider's multi-regional setup described above with their own multi-regional/global load balancing. In this approach, an application owner builds their own multi-regional load balancer using Cloud compute resources. In addition, this multi-regional load balancer must have higher availability than the target application. Such a load balancer can be placed to serve internet traffic, and can also be used for service to service communication as well. A multi-regional load balancer needs to know the health of the stack across all of the regions and the capacity available in each region so as not to overload other regions with too much traffic. Building a multi-regional highly available Load Balancer requires specialized engineering skills and specialized processes to provide what is essentially described in the Global Services Stack in Section 5.3 (built and managed by Cloud providers).

This deployment model depicted in Figure 6 increases the availability of a multi-regional deployment, but puts the responsibility on the application owners for solving the hard problem of connecting regions in a way to optimize for lower latency, resource utilization and health. While sophisticated application owners build additional layers on top of Cloud providers, it is not expected that all application owners, who want global ubiquitous applications, will build their own traffic management, but rather use the global deployment archetypes described next.

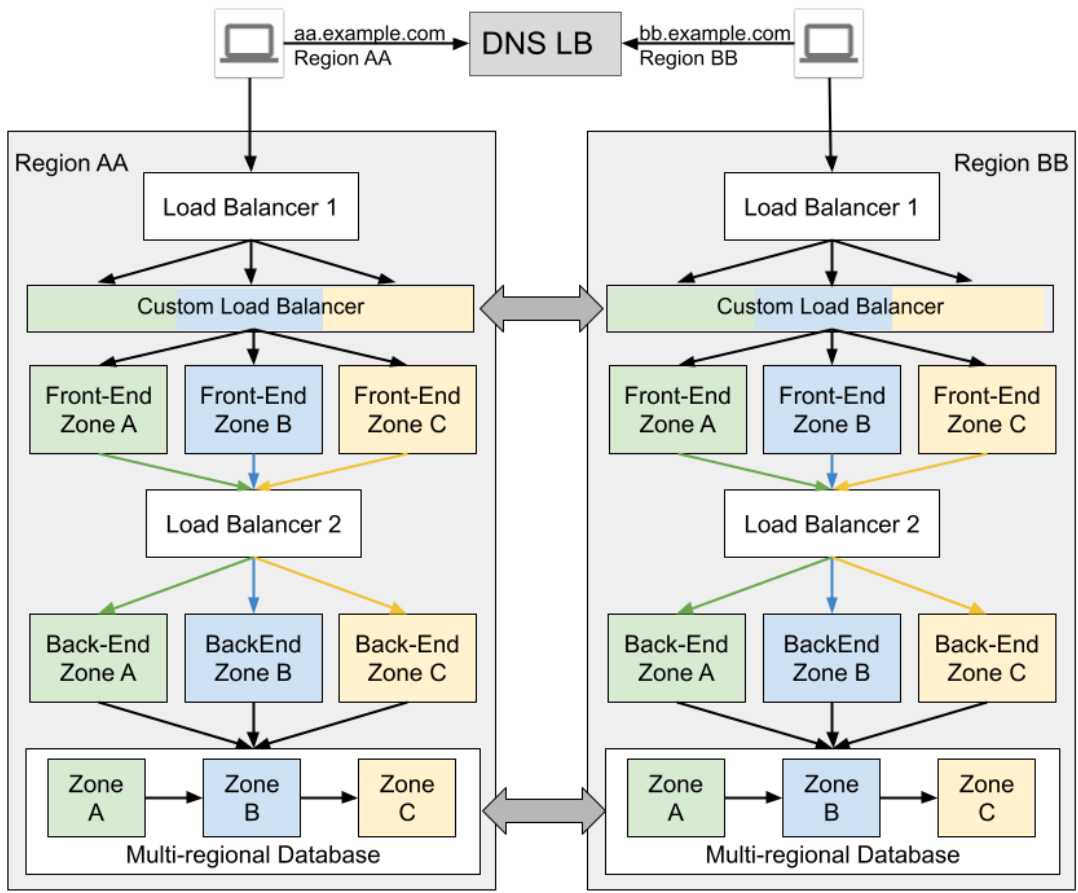

Figure 6: DNS with Custom Multi-Regional Load Balancing deployment model.

## 5. Global

Consumer applications may evolve into global applications due to the global nature of the business, and/or the need to optimize for end-user latency and experience. This means businesses want to serve their cached and dynamic content as close to the users as possible no matter where users are located (both where they live as well as where they travel). In addition, as part of running a global business, application owners have to contend with global events that produce traffic spikes, as well as defend against massive DDoS attacks from around the globe. The difference between multi-regional and global deployments is that while multi-regional creates a deployment from regional building blocks where the application is aware of what region it is running in, a global deployment builds on a globally available fabric of network, data storage and databases that allow the application code to be location unaware.

There are several deployment models for global applications. Below we discuss a few popular ones, but more variations are possible. These models are typically present in the Cloud and not on-premises as they require large global investments in network infrastructure and infrastructure systems that power these applications.

The global model also assumes the data is globally replicated and available in all regions where services run. This is because requests will be load balanced across regions, so multiple regions will need to have access to and read/write the same data. An example of such systems are Google's Spanner [8] and CockroachDB [57]. With global databases like Spanner and CockroachDB, the application can be accessed from any region around the world to perform SQL transactions, with strong consistency and 5 nines availability. Some applications that do not require strong consistency due to their business requirements may achieve five nines with asynchronously replicated and eventually consistent data systems.

### 5.1 Global Anycast

The next step in the evolution of application traffic serving is using Global Anycast as an alternative to DNS Load Balancing to create a deployment capable of instantaneous failover of internet traffic should a multi-regional application become unavailable in one of its regions.

Global Anycast uses a single IP for the application to route traffic from a sender to the topologically closest destination IP address for a group of potential receivers, which for Cloud providers means edge load balancers. Google announces IPs via the Border Gateway Protocol (BGP) from multiple points across its global network [53]. A deployment model that uses Global Anycast eliminates the need for DNS Load balancing with multiple domain VIPs, since the application only needs a single Global Anycast VIP, but it still needs to address the following two issues:

- Too many close-by users can overwhelm an edge site where the traffic is being sent to.
- BGP route calculation might reset connections because of "route flap" [55], which happens when there is a pattern of repeated route withdrawal and re-announcement. This can happen because of frequent problems on a particular link or misconfiguration or mismanagement of routers.

To address these issues, Google developed stabilized anycast using the Maglev [13] network load balancer. This solved the problem of route flap by redirecting a flapped request to a peer Maglev that is responsible for the connection [6]. Maglevs are deployed in each edge location and if an edge location goes down, BGP will reroute to Maglevs in the next closest edge location. In addition, Google uses global load balancing for the Global Anycast LB itself to distribute traffic to edge sites to ensure an edge site is not overloaded. The algorithm considers incoming Requests Per Second (RPS) load and the capacity of edge proxies and assigns new connections to each edge to ensure the best utilization of edge proxies, while at the same time optimizing for user latency.

Google Cloud uses stabilized Anycast technology as a front to the Cloud HTTP(S) Load Balancer [30]. This load balancer provides application owners with the ability to have a single global VIP that represents their global application deployed anywhere in the world.

**5.2 Global Anycast LB with Isolated Regional Stacks**

In this deployment model, a Global anycast LB ingests traffic and then sends traffic to the regional LB in the region containing the application owner's compute resources, depending on geo-mapping, health and weights. Other Cloud providers have also developed products using Anycast - Azure Front Door [46] and AWS Global Accelerator [3].

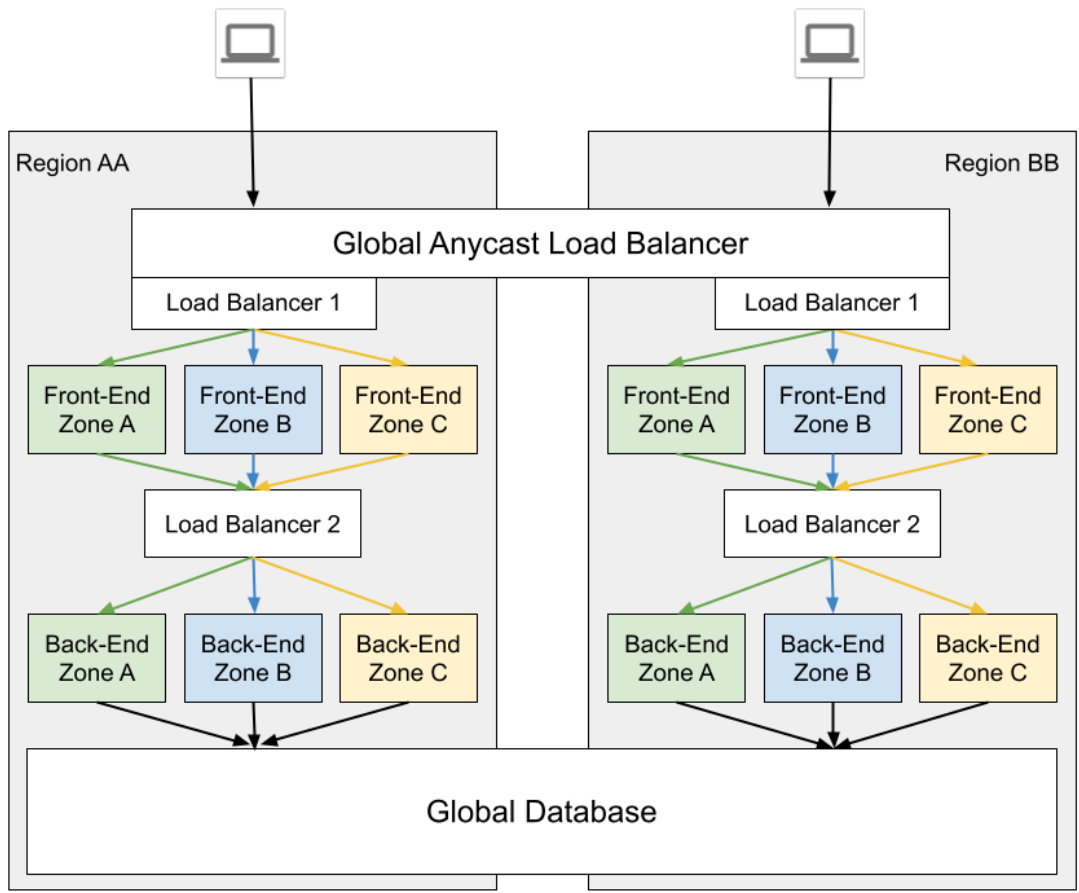

Figure 7: Global Anycast with regional isolated stacks and global database deployment model.

This approach, depicted in Figure 7, uses a regional stack like the prior one, and replaces the DNS LB with a Global Anycast LB. This means the application still has to build up an understanding of the health of all layers within a given region and provide this to the Global Anycast LB. This is the same as described in the DNS LB Section 4.3, where the Global Anycast LB only understands the health being propagated up to it via the Load Balancer 1 layer from each region.

The advantages of a Global Anycast LB are:

- Using DNS with Global Anycast means DNS always resolves a domain to the same single VIP. Lack of reliance on specific DNS resolution means that load balancing between the regions will be done instantaneously by Maglevs and not subject to DNS TTL.
- A single global VIP simplifies an application owner's setup as there is no need to use DNS LB and manage multiple IPs. Using Global Anycast with DNS will resolve a domain to the single global VIP anywhere in the world and traffic will reach the closest healthy destination with available capacity.

The disadvantages of this approach are:

- Since this is still a regional-based stack, the application still needs to implement propagation or aggregation of health across services in regional stacks.
- The application owner does not have control over traffic distribution from clients to the edge location where the LB service resides.  In comparison, with DNS LB the application owner could redirect traffic administratively if needed. For example, if an application itself has a partial issue in a given region, but the health checks are passing (i.e., gray failure), with DNS it is easy to tell it to not send any traffic to the VIP having the issue for that region.

This deployment architecture is for global user-facing applications.

### 5.3 Global Services Stack

In this deployment, services are global. The data is also global and synchronously or asynchronously replicated and available in all regions where services run (e.g., using Google Spanner [8]). In addition, having a global network is important to making a Global Services Stack possible.

This application deployment is targeted towards applications with a worldwide audience, that receive traffic spikes, serve a large volume of traffic, must run economically and need five nines availability. These are typically large-scale global applications deployed over three or more regions and a large number of microservices (a hundred or more) all communicating with each other, with global load balancers between them. At a high level, this approach puts a Global Load Balancer in front of each microservice. Given the large number of microservices, there is typically distributed ownership with each team owning one or a few microservices [39]. Having each microservice (or set of them) having their own Global LB provides the ability to manage traffic and reason about each microservice independently, which fits well with the distributed ownership of the microservices that make up the overall application.

This global load balancer provides the global service-to-service communication for each microservice in the stack. This functionality is provided by either middle proxies or by using a global service mesh with sidecar proxies or even a proxyless gRPC service mesh [52]. Using a managed service mesh has an advantage in that it aids in managing tens to hundreds of microservices with integrated load balancing, health checking and autoscaling without needing to take care of proxy resilience and availability.

With a service mesh, the global service-to-service communication supports HTTP(S)/gRPC and TCP/UDP traffic. Global service communication (called east-west or service-to-service load balancing and routing) optimizes traffic globally for each microservice in the stack, so communication at each source-to-destination pair of services exhibits the lowest latency. This makes sure the destination service is not overloaded and redirects traffic to the closest available region in case of failure or administrative maintenance. The placement of services in a Global Services Stack is economical as only one zone is needed in each region where the application wants to run.

Let us consider the example.com application in a Global Services Stack deployment as depicted in Figure 8. A request to example.com is resolved to the Global Anycast VIP and the request is subsequently sent to the Global Front-End Load Balancer to decide where to serve this request. In addition, each path in example.com (assume example.com/video and example.com/photo in this example) is registered with the Global Front-End Load Balancer so they can be routed to different services. Depending on the path chosen (example.com/video or example.com/photo) as well as other routing options, the request is mapped to a different service, in this case a video service or photo service. This allows each path to have separate health checking, which was not achievable when using DNS LB described in Section 4.3. Each service may be deployed independently as a microservice with its own Front-End service, or combined into a shared Front-End service. Health checking for each path is done independently and a shared Front-End service may reply as healthy for video requests and unhealthy for photo requests allowing global load balancing to direct traffic separately for each service. Depending on the health, geographical proximity, and capacity of the Front-End service, requests will be forwarded to the most appropriate instances.
The same approach happens at the Global Back-End Load Balancer when the Front-End service wants to send requests to the Back-End service. For example, assume the Back-End service in region AA is unhealthy. Traffic from Front-End in region AA will be rerouted to the Back-End in region BB or region CC by the Global Back-End Load Balancer, without the need to propagate health up the stack (as described in the DNS Load Balancing with Isolated Stacks deployment model in Section 4.3).  With the Back-End in region AA unhealthy, the Global Back-End Load Balancer will assess the health, geographic proximity, and capacity of each individual microservice/service in all zones of regions BB and CC, and requests will flow to the most appropriate Back-End instances. From there on, the closest, and most likely local, database will be used to read or write the data.

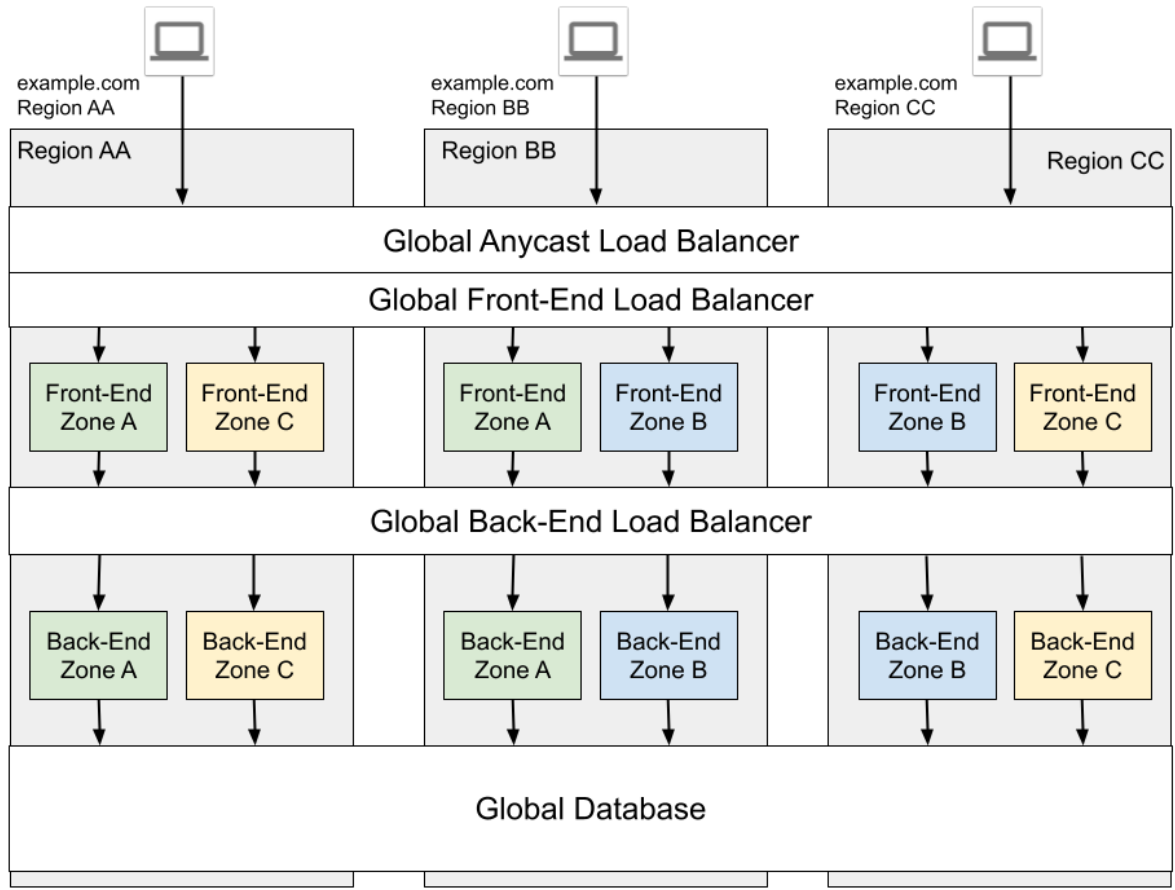

Figure 8: Global Services Stack deployment model.

The following are benefits of the Global Stack Services deployment approach:
- Allows every layer and microservice in the stack to understand the health of and load to that layer in order to load balance for each layer. In comparison, Multi-Region deployments (Section 4), DNS LB with Isolated Stacks (Section 4.3), and Global Anycast LB with Isolated Stacks (Section 5.2) approaches require applications to stitch together stacks and health signals to deal with failures. In addition, all of these approaches lack sufficient understanding of service capacity for load balancing down the stack.
- As shown in the example.com example, a whole region does not become unhealthy for an application if one microservice in the stack becomes unhealthy in that region, compared to the DNS LB approaches.
- Traffic spikes are automatically in real time load-balanced around the world (regions) as needed to keep the application available. Traffic spikes could be caused by users (e.g., triggered by inorganic events) or by services (e.g., an accidentally created DoS attack from lack of exponential backoff).
- This deployment is the cheapest option to run a worldwide application with the highest availability, since capacity can always be efficiently used. In Google Cloud, we integrate global load balancing with autoscaling to allow an application to scale up and down as the load changes [30].

The following is the disadvantage with this approach:

- Global service-to-service communication means the potential for a global outage if the Global LB has an issue. This requires the utmost care in operation of such services. Here are some of the practices we use at Google to manage risk:
    - Rollout of binaries to a global service are done incrementally and done zone by zone. Even within a zone, techniques such as blue-green deployment or rolling updates are employed.
    - A new version of a service workload is canaried for some time in the first zone to understand the impact of a new binary on the availability and latency of the application before resuming rollout to other zones and regions.
    - Changes to the configuration of a global service are done incrementally and applied to one zone or region at a time, or to a percentage of entities that consume configuration (either service workload replicas or collocated networking). A combination of percentage rollout within a zone with per-zone rollout minimizes the blast radius, should the configuration produce a crash in the service's software components.

When using a global stack there is no requirement for services to be global; they can be downscoped to regional when it makes sense. It is up to the particular application architecture, and a mix of regional and global services in one application deployment is possible.

**6. Hybrid**

Hybrid applications that consist of running across a combination of on-premises and cloud are becoming increasingly common. While on-premises applications have grown organically, they still fit into the archetypes and models described in this paper. On-premises software stacks will continue to evolve and be connected with the Cloud, to the point where on-premises can be considered to be another form of a zone or region of a public cloud. Hybrid application availability and resilience can then be improved by creating deployment models that leverage failover between on-premises and Cloud, as well as coordinating the execution of the parts of the application and its services.

There are different deployment models used for these different hybrid applications scenarios, and the below are a few examples:

- Cloud frontend serving for on-premises applications
    - Use case: Better network latency and security for applications.
    - In this scenario, data resides on-premises. Incoming traffic (typically from the internet) is ingested into the Cloud using one of the frontend serving architectures for different archetypes. As traffic is ingested, Cloud-managed services such as content distribution networks, DDoS protection, or access policies are applied and enforced. Then the traffic is sent to the on-premises deployment for further processing.
- Cloud disaster recovery for on-premises applications
    - Use case: Backup of important data for redundancy.
    - Some application owners prefer to use on-premises deployment and sync their data to the Cloud for recovery in case something happens with their on-premises data. In this scenario, the data placement archetype (zonal, regional, multi-regional, or global) can be employed, depending on the Cloud provider's availability and the application owner's needs.
- Replicated application between on-premises and Cloud
    - Use case: During migration from on-premises to Cloud or when traffic demand can grow inorganically.
    - Incoming traffic (typically from the internet) is ingested into the cloud using one of the frontend serving architectures for different archetypes. The application is replicated between cloud region(s) and on-premises datacenter(s). Data is also replicated between cloud and on-premises as if it were cross-region replication between cloud regions. Traffic management capabilities allow for scenarios such as:

- Traffic can be directed to the closest regional application stack, independently of whether the stacks are in the Cloud or on-premises.
- Allow the balancing of application capacity between the Cloud and on-premises parts of the application.
- Burst into Cloud when on-premises application capacity is exhausted.
- Failover to boost overall application availability when on-premises application is unhealthy.
  - Data replication is also being supported between Cloud and on-premises. For example, a database can be replicated across the Cloud and on-premises, as with Cloud SQL [26].
- Using first-party Cloud services with an on-premises application
  - Use case: The application needs specialized services that are easily available in the Cloud and hard to get on-premises.
  - In this scenario, the application is mixed, with some services residing on-premises and some on public Cloud. Services residing in the Cloud are potentially managed first-party services, such as Google Cloud BigQuery and Machine Learning.

**7. Multi-Cloud**

To improve availability even further, the application may be deployed across multiple public clouds, but there are many areas of investment that need to be made to make this an accessible option for businesses.

A multi-cloud application has the potential for the highest availability, as it removes reliance on a single Cloud provider to be always up, and provides cross-cloud load balancing for traffic spikes. In addition, using multiple clouds provides more vendor optionality and choices for the application owner. Also a multi-cloud application may have lower latencies as it gives more options to distribute the load in a given geography across cloud providers.

The current state of this deployment is in its infancy and some of its challenges include:

- **APIs** - a multi-cloud application needs common APIs to effectively run its application across clouds. There are many areas where the APIs are similar enough across clouds (e.g., Object/Blob storage), where others are fairly different. Increasingly there are a number of open source API standards that are becoming common across clouds (e.g., Kubernetes APIs [42], Envoy Proxy APIs [15], and Istio [34]), which will aid multi-cloud deployments. Another approach some applications have taken is to leverage the base semantics that are common across the clouds and abstract the cloud-specific differences away into a client library layer the application uses. The library layer can translate the requests to the appropriate APIs depending on the cloud being used.
- **Operational Complexity** - to run multi-cloud applications, multi-cloud operational tools and orchestrators for binary configuration, upgrade, rollout, monitoring and debugging must be developed. The complexity of building such tools is based on the challenges of not having common cross-cloud APIs as well as the operational differences in platform. To help address this, application owners can use a cross-cloud application management platform such as Anthos to provide consistent development and operational experiences across clouds. In addition, application owners can use cross-cloud configuration tools such as Terraform to simplify configuration to provision resources in multiple clouds [9].
- **Load balancing** - there are multiple ways of load balancing between public and private clouds. The majority of them are only in the beginning of a multi-cloud journey and this is an interesting space to watch:
  - **DNS LB** - similar to the DNS LB with Isolated Stacks deployment described in Section 4.3, a multi-cloud application could use DNS LB to route traffic across clouds.
  - **Client-driven traffic routing** - each client receives IPs/VIPs for each of the clouds and distributes traffic based on either routing or load balancing configuration. Configuration is delivered via a control plane. An example of this approach is Envoy Mobile [16].
  - **Global Anycast LB with isolated stacks across clouds** - in this deployment, one cloud has the primary Global LB that distributes traffic between primary and secondary clouds, based on algorithms chosen by the application owner. This approach assumes network connectivity and that the global load balancer has the endpoints from other clouds registered with it. The distribution algorithms may be (a) geographically optimized (e.g., send to the closest cloud), (b) burst to the other cloud due to lack of capacity, which is typically from private to public cloud, and (c) failover to the other cloud based on the health of the service behind the Global LB.

- ○ **Global Services Stack** - this is similar to a single-cloud Global Services Stack deployment described in Section 5.3, where each layer either operates across clouds or the serving stack is distributed across clouds. In this case, a given service is replicated across multiple clouds and traffic is distributed based on geography or other constraints. This deployment model is hard to achieve today, and for this approach, having common cross-cloud APIs is important.

- **Networking and Security -** A multi-cloud application can have its services in different clouds connected to one another via public (Internet) or private (interconnects and VPN tunnels) connections. Connectivity is subject to QoS and bandwidth management, and network level encryption via IPSEC or SSL/TLS. For private connectivity, underlying pipes are shared between applications, and cloud providers must manage resources shared between different services and their customers.

  Multiple networking topologies [20] are supported for public and private connectivity based on the needs of the application:

  - ○ **Flat network** - In this model a flat network spans multiple clouds and all services can communicate with one another. Firewalls are enforced on both sides. In addition, zero trust or end-to-end application security via mTLS authentication and authorization between services in different clouds is used. This requires either homogeneity of compute environments such as Kubernetes and hence the same credentials, or have federated identities for heterogeneous compute environments.

  - ○ **Gateway model** (ingress and egress) - In this model, there are gateways on both sides. Depending on the direction of the traffic, it is either ingress or egress gateways. Ingress gateway protects access to the services by providing a security perimeter to enforce access to the limited number of services by allowed or denied roles or IP addresses. The access is enforced for "workforce" (people) and "workloads" (services). Egress gateway protects source workloads and data, from events such as exfiltration attempts, and also enforces policies for destination services in other clouds. Zero trust model is also possible here by Gateway terminating mTLS from the client and creating another mTLS connection to the server.

  - ○ **Handover model** - In this model, there is a shared environment between two clouds and data from one cloud is uploaded to this shared environment and picked up by the workload from another cloud (e.g., using Pub/Sub or worker queues). There is no private network connectivity between parts of the application. This model is not used by serving applications, but rather by processing applications and data analytics.

  These topologies apply to hybrid deployment archetype also.

- **Data management** - Multi-cloud database [40] and storage solutions that replicate across clouds [38] would need to be used. Approaches can be considered where the primary for the database is in one cloud, and read replicas in another cloud [48]. If the primary cloud has issues, then first try to failover within that cloud, otherwise failover to another cloud.

- **Cost** - A multi-cloud deployment may entail higher costs, with a tradeoff for higher availability and cross-cloud optionality. A few examples are:

  - ○ Data duplicated across clouds will typically be stored also redundantly within each cloud, which costs more than just keeping the data replicated within a single cloud.

  - ○ The egress costs for replicating and duplicating data across clouds may be higher than just storing the data replicated and durable within a single cloud.

  - ○ There can be resource inefficiency unless the cross-cloud load balancer fully understands the capacity utilization in each cloud. This includes understanding the actual capacity and load to the microservice layers in each cloud.

The multi-cloud approach is promising for increasing availability and optionality, but there are many areas of investment that need to be made to make this an accessible option for businesses.

**8. Comparison**

Throughout this paper, we examined how to span zones, regions, and geographical reach to achieve availability for serving applications. To achieve higher availability, an application deployment has copies of its serving stack, using either (a) an additional failover copy or (b) additional active serving stacks to load balance across. As the application increases its geographical spread from zonal, to regional, to multi-regional, to global, a higher number of nines of reliability can be achieved. The deployments that achieve the highest availability are multi-regional, global, and multi-cloud.

Multi-regional and global deployment archetypes need databases, object stores, and data caches that provide access to shared state. Depending on the type of datastore needed, there is a spectrum of multi-regional and global data store deployments, using either synchronously or asynchronously replicated data, to choose from. The type of datastore used depends on the application's access patterns and type of data that it is serving, whether it serves read-only cached data (e.g., songs, images, or directions), can function well with eventual consistency (e.g., putting items in an online shopping cart), or requires strong consistency. In addition, the application owner needs to determine how much redundancy and freshness of data is needed for business continuity.

Tables 1 through 5 provide a comparison between the different deployment archetypes and models. Some deployment archetypes (table for each archetype) have more than one deployment model (first column). We compare each archetype and deployment model pair based on characteristics that are used to judge the risks of application failures. The second column is the potential scope of failure, which we separate into four types: zone, region, global and cloud (for multi-cloud deployment). The third column describes, at an abstract level, the type of failure that could cause the corresponding scope of failure to occur.

Therefore, each row in the table describes the application impact and recovery for a given deployment archetype and model assuming the specified scope and type of failure has occurred. Then, for application impact and recovery, the last three columns describe: the impact to the application's serving stack's availability (fourth column) , the type of mitigation needed to restore availability because of the failure (fifth column), and whether the mitigation provides instantaneous recovery or not (sixth column). For the Application Down column, "Yes" means the application is down with an outage until the specified Mitigation happens, and "No" means the application continues to serve traffic (typically due to load balancing across the scope of failure or after failover completes).

A regional application may experience zonal and regional failures, whereas a multi-regional application may experience zonal, regional and global failures. In addition, the failures could be cascading from smaller scopes to a larger one. For example, cascading zonal failures limited within a region would be considered a regional failure and spreading beyond the region would be considered a multi-regional or global failure.

When we examine What Failed, it could be the cloud service provider, or could be a service in the application, which may include dependencies on third party services. It is important to capture and understand all of the dependencies, their criticality to availability and their risks when considering what can fail and how.

Table 1: Zonal Deployment Archetype Comparison of Risks

| Deployment Model | Scope of Failure | What Failed | Application Down | Mitigation | Instantaneous Recovery |
|---|---|---|---|---|---|
| Single Zone | zone | Zonal infra or managed services | Yes | Wait until zone is back or rebuild app in new zone | No |
| Single Zone with Failover (Figure 1) | zone | Zonal infra or managed services | Yes (during failover) No (after failover) | Continue operation via failover to standby zone | No |
| | region | Regional infra or managed services | Yes | Wait until region is back or rebuild app in new region | No |

Table 2: Regional Deployment Archetype Comparison of Risks

| Deployment Model | Scope of Failure | What Failed | Application Down | Mitigation | Instantaneous Recovery |
|---|---|---|---|---|---|
| Single Region (Figure 2) | zone | Zonal infra or managed services | No | Continue operation from remaining zones in the region | Yes |
| | region | Regional infra or managed services | Yes | Wait until region is back or rebuild app in new region | No |
| Single Region with Failover (Figure 3) | zone | Zonal infra or managed services | No | Continue operation from remaining zones in the primary region | Yes |
| | region | Regional infra or managed services | Yes (during failover) No (after failover) | Continue operation via failover to standby region | No |
| | global | DNS LB (if DNS is used for failover) | Yes for new or expired TTL clients | Wait until DNS LB is back | No |

Table 3: Multi-Regional Deployment Archetype Comparison of Risks

| Deployment Model | Scope of Failure | What Failed | Application Down | Mitigation | Instantaneous Recovery |
|---|---|---|---|---|---|
| Multi-Regional Fully Isolated with Data Sharding (No Failover to Standby) (Figure 4) | zone | Zonal infra or managed services | No | Continue operation from remaining zones in the region | Yes |
| | region | Regional infra or managed services for a single region | Yes for users in region affected, since users are sharded across regions | Wait until shard is back | No |
| | global | DNS | Yes for new or expired TTL clients | Wait until DNS is back | No |
| Multi-Regional Fully Isolated with Data Sharding with Failover to Standby (Figure 4) | region | Regional infra or managed services | Yes (during failover) No (after failover) | Continue operation via failover to standby for region affected | No |
| Multi-Regional with DNS LB (Figure 5) | zone | Zonal infra or managed services | No | Continue operation from remaining healthy zones across the regions | Yes |
| | region | Regional infra or managed services | No | Continue operation from remaining regions | No (DNS TTL) |
| | region | Single app service down | No | Load balance traffic away from the affected regional stack to another region | No (DNS TTL) |
| | global | DNS LB | Yes for new or expired TTL clients | Wait until DNS LB is back | No (DNS TTL) |
| | | Multi-Regional Database | Yes | Wait until database is back | No |
| Multi-Regional with DNS LB & Custom Multi-Regional LB (Figure 6) | global | DNS LB | Yes for new or expired TTL clients | Wait until DNS LB is back | No (DNS TTL) |
| | | Custom LB | Yes | Mitigate custom LB failures | No |
| | | Multi-Regional Database | Yes | Wait until database is back | No |

Table 4: Global Deployment Archetype Comparison of Risks

| Deployment Model | Scope of Failure | What Failed | Application Down | Mitigation | Instantaneous Recovery |
|---|---|---|---|---|---|
| Global Anycast with Isolated Stacks and Global Database (Figure 7) | zone | Zonal infra or managed services | No | Continue operation from remaining healthy zones across the regions | Yes |
| | region | Regional infra or managed services down | No | Continue operation from remaining regions | Yes |
| | region | One instance of application microservice in region down | No | Continue operation of this microservice from remaining regions | Yes |
| | global | Global Anycast | Yes | Wait until Global Anycast recovered | No (potentially have backup VIP) |
| | | Global Database | Yes | Wait until database is back | No |
| Global Services Stack (Figure 8) | zone | Zonal infra or managed services | No | Continue operation from remaining healthy zones across the regions | Yes |
| | region | Regional infra or managed services | No | Continue operation from remaining regions | Yes |
| | region | One instance of application microservice in region down | No | Region(s) not considered down. Load balance individual microservice to another region | Yes |
| | global | Global Anycast | Yes | Wait until global Anycast recovered | No (potentially have backup VIP) |
| | | Global Database | Yes | Wait until database is back | No |
| | | Global service to service LB (gateway/middle proxy) | Yes | Wait until Global LBs are back | No |
| | | Global service to service (proxyless or sidecar LB) | No | Continue in degraded using old endpoints and health | Yes |

Table 5: Multi-Cloud Deployment Archetype Comparison of Risks

| Deployment Model | Scope of Failure | What Failed | Application Down | Mitigation | Instantaneous Recovery |
|---|---|---|---|---|---|
| Client Side Load Balancing | one cloud | Global or multi-regional issue | No | Continue operation from remaining cloud | Yes (if using client-side load balancing) |
| DNS Load Balancing | one cloud | Global or multi-regional issue | No | Continue operation from remaining cloud | No (DNS TTL) |

**8.1 Failover-to-Standby vs. Load Balancing**

When comparing the failover-to-standby deployment models to those that use load balancing across failure domains, the failover-to-standby model has the following concerns an application owner needs to address:

- Failover-to-standby models do not provide instantaneous recovery due to DNS TTL delays and therefore display unavailability even if for a short period of time. The startup time of the standby stack and the ability for it to go from zero load to full throttle is important, as traffic does not trickle in slowly on failover, but rather moves over all at once. For example, when the standby stack takes on primary traffic, its service caches may be cold and there is going to be observed latency impact, as well as additional compute capacity may be needed to process requests as the caches warm up. These failover startup delays can impact availability, and can expose startup issues by putting high load on infrequently stressed code paths that themselves may cause application outages or delays during failover.
- A standby stack must be maintained even though it is unused the majority of the time. In addition, there is always a risk that the standby stack will not be ready to take on primary responsibility during an outage event. To avoid this risk, failover systems should be regularly tested by promoting the standby stack to the primary stack. Special functionality testing, load testing and fault injection tools and monitoring must be used to have high confidence in a standby stack.
- Failover-to-standby models are typically more expensive, as the extra standby serving stack must be up and running, instead of reusing active additional setups in other zones or regions via load balancing.

In comparison, an application which is using a load balancing-based deployment model steers traffic away from the failure domain having an issue. This means the application does not have to deal with the above issues because every zone and region is always taking traffic. However, it does require the application to either provision for N+1, where 1 is a failure domain and N is needed to serve traffic for the application, or use auto-scaling with a large enough buffer. The failover in load balancing is triggered by the change in the health-checking results or by detection of outlier instances that appeared unhealthy. Typically, there is a threshold of unhealthy endpoints in a failure domain configured, so once a number of unhealthy endpoints crosses the threshold, traffic starts to be load-balanced away from the failure domain.

Some cloud provider load-balancing products implement the notion of gentle failover, where instead of moving traffic over in one step, the traffic trickles over to the other zones and regions to warm up the caches. Even though all locations have warmed up caches for their standard traffic patterns, when traffic is load-balanced to a new region, there may still be warming up that needs to occur if this type of traffic is new to that region (e.g., if traffic is language-specific).

**8.2 Regional vs. Global Application Stacks**

It is also interesting to compare regional application stacks to global application stacks. With a regional application stack, the whole region has to be pronounced unhealthy if there is an issue with any part of the application in a region, and that region is removed from serving traffic. This occurs even if it is only one service in the application stack that is unhealthy for that region.

In comparison, a global application stack is not considered unhealthy if an individual service in a region has an issue, because each individual service in the application can be independently load-balanced to another region as needed. Independence of individual service or microservice failover makes the Global Services Stack deployment model to be the most cost-optimized and have lower complexity as there is no need to aggregate failures together or propagate them up the stack. This model also matches scaling the management of the running application where different microservices are owned and operated by different teams.

In addition, with the Global Service Stack it is more nimble to expand the service into additional zones and regions, should the business expand into new geographies. For example, some applications may need part of the stack to be as close to the consumer as possible and therefore want to serve only part of the stack from edge locations, where the amount of compute resources and network bandwidth is limited. Because each service and microservice is individually deployed in a Global Services Stack, this provides the ability to deploy only latency-critical services at the edge and leave the rest of the stack in cloud provider regions.

Ownership and management of infrastructure, of the outage when infrastructure fails, and the subsequent recovery is an important dimension to consider. When an application owner needs a global deployment, but only has access to or decides to use regional cloud infrastructure, this requires the application owner to build custom infrastructure that stitches regions together to create a seamless global deployment. This custom global infrastructure is complex to build and operate to achieve high availability, and the ownership falls on the application owner. Using a Global Services Stack deployment provided by a cloud provider moves the complexity and responsibility of custom LBs from application owner to the cloud provider, along with the responsibility for the availability of all the infrastructure needed to provide the global serving stack.

**8.3 Instantaneous Recovery**

The final column in Table 1 examines which cases support instantaneous recovery, which is the ability to switch away from the failed application stack in a zone or region without incurring a delay in sending traffic to an available zone or region. Automatic instantaneous recovery via load balancing is needed for applications requiring 5 nines availability, since they cannot have manual mitigation with a budget of less than 5.25 minutes of unavailability per year. To achieve instantaneous recovery, cloud-native Global Anycast Load Balancing products are preferable to DNS load balancing to avoid delays that the DNS protocol introduces via TTL configuration and clients (e.g., web browsers) disobeying TTLs, as well as due to Global Anycast Load Balancing having integrated health checking. Having said that, when using Global Anycast, the availability of the application is highly dependent on the availability of Global Anycast and its single VIP. This is why there is significant investment into reliability by Cloud providers for Global Anycast load balancing. In addition to the continued investment in reliability, cloud providers are looking at providing a backup VIP to the Global Anycast VIP.

**8.4 Cost of Availability**

The final dimension we want to examine is the cost of availability. When comparing all of these models, every application defines an acceptable probability of failure that makes sense for the business. Based on the tolerance for failure, application and data needs, a deployment archetype is chosen. Each archetype has a cost to achieve the desired number of nines. The higher the desired number of nines, the higher the cost to achieve it. There are multiple dimensions of cost, ranging from the number of instances per zone, number of zones to use, network bandwidth cost for cross-zonal or cross-regional failover traffic, the cost of synchronous and asynchronous data replication, the cost of redundant storage, the cost of the software complexity required to assemble an application within each archetype, the cost of training and professional skills, and the cost of managing the application.

Zonal and regional archetypes are similar in cost, while deploying an application beyond one region can add cross-regional network bandwidth costs. Multi-regional and Global archetypes have even higher cost due to data replication around the world. Hybrid and Multi-cloud deployments can be even more costly due to egress cost, and data storage and replication costs, as well as the cost of redundant compute instances that are not exactly the same in each cloud and on-premises environment, which can result in unutilized resources. For example, the hardware configurations, memory sizes, networking bandwidths, storage latencies, and more, can be different across clouds and on-premises, and need to be optimized differently to arrive at efficient deployments across them. How far the application owner pushes towards using higher availability deployment models comes at a cost, and is a tradeoff based on what the application needs and the impact on application users and the business when there is an outage.

When comparing economics of redundancy in regional and global application stacks, we can observe that global deployments are cheaper, while providing higher reliability. Let us look at an example. Multi-regional deployments typically replicate within a region and run 3 zones for each region for redundancy with the cost of 3*N, where N is the number of regions. This model is typically used across 2 regions. As you increase the number of regions, the cost goes up dramatically, and it makes sense to move to a global deployment from a cost perspective. A global deployment requires only one fault domain (zone) within a region, so the cost is N instead of 3N for the application serving stack. As more regions are added, the costs of additional regions is incremental for a global deployment.

In this section, we compared archetypes and deployment models on geographic redundancy, failover to standby versus load balancing, instantaneous recovery, ownership of the infrastructure, and cost. There is no perfect deployment model, and every application owner needs to pick deployment model(s) that are best for their application, customers and business.

## 9. Related Work

**How to build cloud applications:** The architecture used for cloud applications has evolved from monolith into service oriented architecture (SOA) [41]. SOA results in a distributed system[56] with large numbers of moving parts of services and microservices, each of which has a deployment mode. An application, which is a collection of services and microservices, also has a deployment archetype that needs to be optimized for high availability. [67] examines in detail cloud application survivability concepts such as fault-tolerance, reliability, and availability. They consider failures of software components, hardware infrastructure, and a failure of the zone or one of more regions, as events contributing to cloud application failure and they describe deployment archetypes to sustain these failures. In comparison, our paper examines the deployment archetypes focused on availability, latency and geographical goals across zones, regions and global deployment models. This is just as important to the understanding of the overall architecture as is choosing the type of infrastructure and services (e.g., VMs, containers, serverless, microservices) to use to build your application, which the prior work focuses on. For e-Commerce sites, streaming video services, news sites, gaming platforms, and IoT services that require 24/7 uptime, a multi-regional architecture is preferred [18]. Estrin surveys 3 public cloud providers - Amazon Web Services, Microsoft Azure and Google Cloud in the areas of DNS, CDN, DDoS, storage and databases, while also taking into account cost and operational aspects [18]. We build on this work to examine deployment archetypes beyond multi-regional.

**Traffic Management and Load Balancing for cloud applications**: Each of the deployment archetypes described in this paper requires traffic distribution, load balancing and high fidelity health checking to meet their needs. The approach needs to make sure geographically distributed services and microservices are not overloaded and the application as a whole is optimized for latency. The higher the expected availability level, the more traffic management support needed, which includes load balancing, health-checking, autoscaling, health-driven failover, cross-regional overflows, rate limiting and load shedding. The approaches described in our paper are standard ones existing in industry, and we have focused on which ones to use for the different archetype models described. The following is a set of related work that provides the details for how these approaches work for DNS, Anycast and healthchecking.

[4] described using DNS Load Balancing to distribute internet traffic to the closest geographically as well as healthy cloud application regional stacks. Such distribution optimizes dynamic application capacity and instant health status of geographically distributed instances. [44] describes how DNS Load Balancing evolved as a traffic management tool for cross regional traffic management.

An important part of load balancing-as-a-solution is health-checking. [50] describes how Fastly built a distributed healthchecking system for making a decision about optimal traffic distribution and failover should geo distributed servers become unhealthy. Anycast Load Balancing [66] described anycasting communication paradigm at the application layer as a way to find latency-optimized replicated servers by having an anycast resolver to map an anycast domain name and a selection criteria into an IP address.

Tenereillo [58] showed that while DNS is often used for global server load balancing [32], it is not suitable for global cloud applications that require five nines of availability because of inherent delays built into DNS protocol. Globally ubiquitous applications that often use CDN [19] cannot afford DNS slow reaction during failover. Instead they use anycast for geographic routing of internet traffic to their application stacks that are spread out across regions [36]. Google Cloud [30], Microsoft Azure [46] and AWS [3] built global anycast products for global applications to address this.

**Hybrid applications**: An application that is split between on-premise data centers and a public cloud faces the pressure of how to evolve the application parts that reside on-premises to work with cloud as described in [1, 47, 57]. An approach often taken is to evolve the application to a service oriented architecture through partial, in-place re-architecture and rewrite, or building the bridge to service-based application deployment [29]. We view on-premises datacenters evolving into a private region in a multi-regional deployment. The hybrid application is then built using desired multi-regional or global deployment archetypes described in this paper.

**Multi-cloud applications**: A multi-cloud application can potentially achieve the benefit of increased availability, and optimized latency [63] due to using separate cloud stacks, that do not share failures due to software bugs, and wider geographical reach and presence. This cannot be achieved without application services using cross-cloud APIs and portable open source solutions as discussed in [12, 49] to assure that the parts of application can be either replicated or interconnected across the clouds.

## 10. Summary

In this paper, we examined several deployment archetypes for Cloud applications. The deployment archetypes are evolutionary, from zonal, to regional, to multi-regional, to global, to hybrid, and to multi-cloud, where each step progressively provides increasing higher availability and better end-user latency.

Each archetype has its place and importance according to unique application requirements and the tradeoffs it provides. Applications will want to push as far up the deployment archetypes as possible. For user facing-applications, a DNS Load Balancing with separate stacks is the standard and familiar approach for multi-regional and cross-cloud deployments.

For applications that want to achieve the highest level of availability, the Global Services Stack deployment model is preferable to stitching isolated regional service stacks together with DNS Load Balancing, due to its ability to load-balance with fully integrated health monitoring, and understanding capacity management. Multi-cloud with load balancing is the deployment model to watch as it evolves, with related technologies also rapidly evolving. This model drives the importance of open source APIs and client-side traffic management.

**Acknowledgements**

We would like to thank Andi Gutmans, Geoff Voelker, Amit Ganesh, Kara Moscoe, Sachin Gupta, Ben Treynor, John Laham, Sam Greenfield, Chris Taylor, Dave Nettleton, Nirav Mehta, Olaf Schnapauff, Ines Evid, Davis Hart, Michael Abd-El-Malek, Sameet Agarwal, Zach Seils, Jai Haridas, James Duncan, Barbara Stanley, and Philippe Poutonnet for providing valuable feedback on this paper, and everyone working on Cloud whose work informed and inspired this survey.